# A BiLSTM-CNN based Multitask Learning Approach for Fiber Fault Diagnosis

**Khouloud Abdelli[1,3], Helmut Grießer[1], Carsten Tropschug[2], and Stephan Pachnicke[3]**

[1]*ADVA Optical Networking SE, Fraunhoferstr. 9a, 82152 Munich/Martinsried, Germany*
[2] *ADVA Optical Networking SE, Märzenquelle 1-3, 98617 Meiningen, Germany*
[3] *Christian-Albrechts-Universität zu Kiel, Kaiserstr. 2, 24143 Kiel, Germany*
*E-mail: KAbdelli@adva.com*

**Abstract:** A novel multitask learning approach based on stacked bidirectional long short-term memory (BiLSTM) networks and convolutional neural networks (CNN) for detecting, locating, characterizing, and identifying fiber faults is proposed. It outperforms conventionally employed techniques. © 2021 The Authors

## 1. Introduction

Optical time reflectometry (OTDR), a technique based on Rayleigh backscattering, has been widely applied for fiber characteristics' measurements and for fiber fault detection and localization. However, OTDR traces can be hard to interpret even by experienced field crews. Fiber fault real-time detection is considered as the industry standard to ensure optical network survivability and reliability. Therefore, it is crucial to develop an automated reliable technique capable of accurately and quickly detecting, localizing, and identifying faults given noisy OTDR data and thereby reducing operation-and-maintenance expenses (OPEX). In the past several OTDR event analysis approaches have been proposed. Conventionally, the OTDR event detection technique relies on a two-point method combined with least square approximation, which is coarse and noise sensitive. Recently, data-driven approaches based on Long Short-Term Memory (LSTM) have been proven to efficiently solve fault diagnosis problems given sequential data [1-2] and could provide an untapped potential to tackle the fiber fault diagnosis problem.

In this paper, we develop a novel hybrid multitask learning framework combining bidirectional long short-term memory (BiLSTM) networks and convolutional neural networks (CNN) to diagnose optical fiber faults when analyzing noisy OTDR data. The multi-task learning architecture is composed of a BiLSTM-CNN shared hidden layer distributing the knowledge across multiple tasks (fault detection $T_1$, fault localization $T_2$, fault characterization $T_3$ and fault cause identification $T_4$) followed by a specific task layer. The proposed approach is applied to noisy experimental OTDR data, whose SNR (signal-to-noise ratio) values vary from 0 dB to 30 dB, comprising multiple fiber faults with various characteristics. The experimental results show that our model (i) achieves better fault diagnosis capability compared to several other ML techniques namely CNN, LSTM, and BiLSTM and (ii) outperforms the conventional OTDR event analysis methods particularly for low SNR values.

## 2. Setup & Configurations

The SNR of the OTDR trace depends on the receiver photodiode sensitivity, the pulse power used in the measurement, the attenuation, the fiber length, and the averaging of repeated measurements or low-pass filtering of the trace. As the fiber is given and receiver sensitivity as well as launch power are limited, the only parameters left to influence the SNR, are the pulse width, the filtering (which allow to trade SNR against location accuracy) and the averaging (which can improve the SNR at the cost of a longer measurement time). Therefore, for our experiments, the SNR is considered as the relevant figure-of-merit and averaging is used to influence it.

### 2.1. Experimental Setups

The experimental setups shown in Fig. 1 are conducted for recording OTDR traces incorporating different types of fiber events with various reflectances and losses. Optical components like connectors, VOAs, reflectors are also used to model specific faults in the fiber link. To generate non-reflective faults, involving only attenuation and no reflection, with small decrease, an angled physical contact (APC) connector is placed at the end of the 1 km fiber depicted in Fig.1 (left). These non-reflective faults are identified as fiber-bend. To produce non-reflective events with abrupt decrease, an APC connector is located at the end of setup 1 (i.e. the end of the 5 km fiber). The generated non-reflective faults are classified in that case as tilted fiber-cut. The reflective faults (Fresnel reflection) with small peaks are induced by placing a reflector at the end of the 1 km fiber shown in Fig. 1 (right). They are identified as longitudinal connector-misalignment. The reflectance of these events is varied. By putting a physical contact (PC) connector at the end of the setup 1, reflective events with sharp peak and abrupt decrease are caused. They are classified in such circumstances as perpendicular fiber-cut. The coupler, the variable optical attenuators (VOA) and the fibers are individually connected with patch cords. Every connection between the aforementioned components is with APC connectors. Some of the connectors are deliberately dirty. As a result, merged (combined) events comprising either two overlapped reflective faults or overlapped non-reflective and reflective events are induced. The attenuation of the non-reflective and combined faults is modified by varying the VOAs settings. The laser power is varied from 7 dBm to 17 dBm.

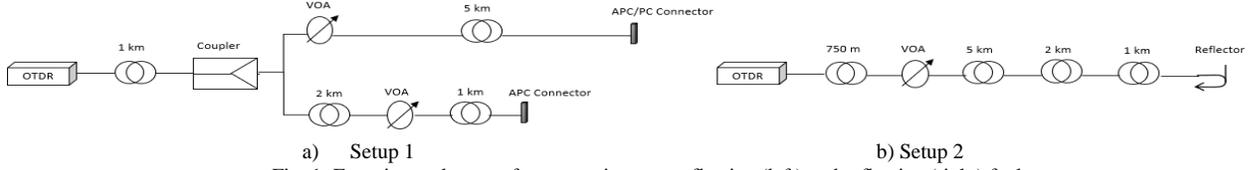

Fig. 1: Experimental setups for generating non-reflective (left) and reflective (right) faults.

Various OTDR records are collected and averaged ranging from 62 to 64,000 traces. The OTDR configuration parameters namely the pulse width, the wavelength and the sampling rate are set to 50 ns, 1650 nm and 8 ns, respectively. Figure 2 depicts an example of an OTDR trace generated using setup 1. It shows the non-reflective event at 4.014 km, the reflective event at 6.012 km and the induced merged events at 995 m and 3.003 km.

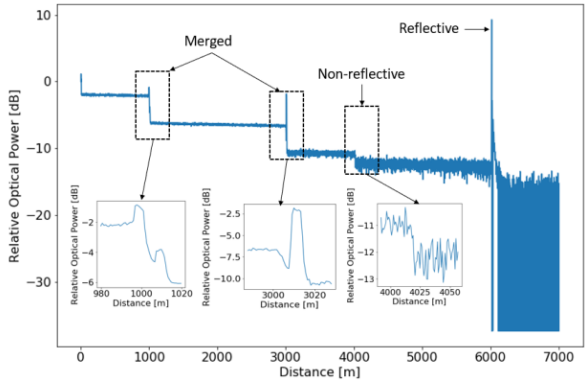

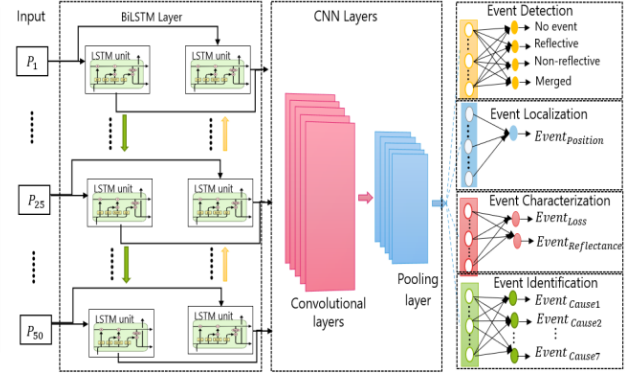

Fig. 2: OTDR trace

Fig. 3: Proposed model architecture

### 2.2. Data Preprocessing

From each OTDR trace generated by setup 1, 5 sequences of length 50 are extracted randomly: one sequence containing no event, two sequences including the merged events, one sequence involving the non-reflective fault and one sequence holding either reflective or non-reflective events depending on the connector type placed at the end of the setup. Whereas for each OTDR trace created using setup 2, two sequences of length 50 (i.e one sequence containing no event and one sequence including the reflective event pattern) are randomly selected. We assigned to each sequence the event type (no event, reflective, non-reflective, merged), the event position index within the sequence, the event loss and/or reflectance and the event cause. The "no event" cause is labeled as no fault (class 0). The "reflective event" causes are classified as either longitudinal connector-misalignment (class 1) or perpendicular cut (class 2) [3]. Whereas the "non-reflective" fault causes are labeled as fiber bend (class 3) or tilted fiber cut (class 4) [3]. The merged event causes (class 5, class 6) are either dirty connector and/or fiber bend. In total, a data set composed of 30,112 samples, whose SNR values vary from 0 to 30 dB, was built and normalized.

### 2.3. Machine Learning Model

Given that the tasks $T_1$, $T_2$, $T_3$, and $T_4$ are highly related and can significantly benefit from the knowledge sharing across them, a multi-task learning based BiLSTM-CNN framework is implemented to learn the tasks simultaneously in order to enhance the generalization capability. As a BiLSTM is well suited to process the OTDR sequential data and to capture the long-term dependency and a CNN is proven to be good at local features extraction, a hybrid deep learning approach by integrating BiLSTM and CNN is chosen as the shared hidden layer of our model to improve the performance by combining their strengths. The architecture of a shared hidden layer consists of one BiLSTM layer with 32 cells followed by CNN layers containing mainly one convolutional layer having 32 filters with the max pooling layer succeeded by a dropout layer to prevent overfitting. The knowledge learned by BiLSTM-CNN is then transferred to the four task-specific layers composed of 16, 20, 32, 40 neurons, respectively. The overall structure of the proposed model is depicted in Fig. 3. The total loss of our model is computed as the weighted sum of the four individual task losses set to 1.5, 0.5, 1.8 and 1, respectively.

### 3. Results and Discussion

We use several metrics including the detection rate (i.e. detection probability) for the $T_1$ evaluation, the root mean square error (RMSE) for $T_2$ and $T_3$ performance assessment, and the receiver operating characteristic (ROC) curves summarizing the trade-off between the true positive and false positive rates for different threshold settings for $T_4$ evaluation. As depicted in Fig. 4, the detection probability of the different events at a false alarm rate (FAR) of 0.01 increases with SNR, and for SNR values higher than 10 dB, it is approaching 1. The RMSE of position estimation decreases with SNR, and it is less than 2 m for SNR values higher than 20 dB. The RMSE of the loss ($RMSE_L$) and reflectance ($RMSE_R$) prediction decreases as SNR increases. The $RMSE_L$ is less than 0.3 dB for SNR values higher than 20 dB whereas the $RMSE_R$ can be as low as 2 dB for higher SNR values. The ROC depicted in Fig. 4 shows that our model could distinguish between the different event causes.

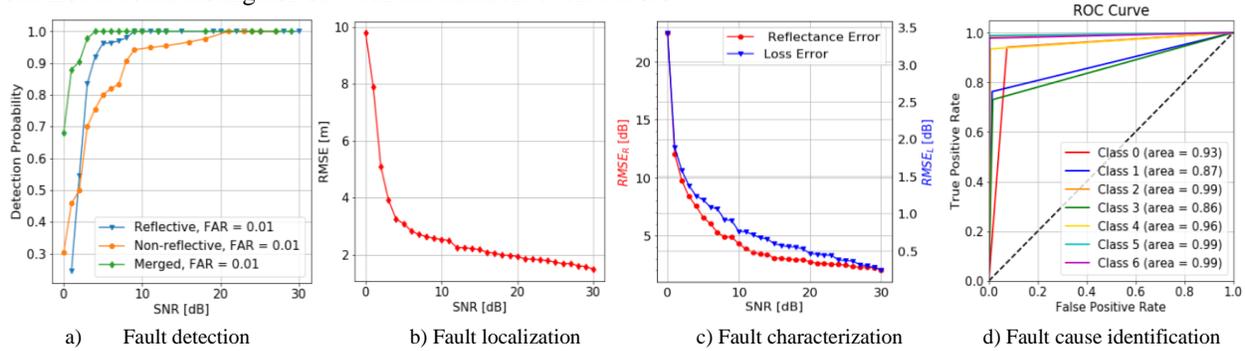

a) Fault detection  b) Fault localization  c) Fault characterization  d) Fault cause identification

Fig. 4: Performance of proposed BiLSTM model

We compare the proposed model with different ML methods, i.e. CNN, LSTM and BiLSTM using different metrics namely accuracy, F1 score (a harmonic mean of precision and recall), RMSE, mean absolute error (MAE), symmetric mean absolute percentage error (SMAPE) and the area under curve (AUC). The results shown in Tab. 1 demonstrate that our method outperforms the aforementioned ML techniques in all metrics.

Table 1. Comparison Results of our model with other ML methods

| Metrics \ Method | CNN | LSTM | BiLSTM | **Proposed Model** |
|---|---|---|---|---|
| $T_1$: Event detection | | | | |
| Accuracy [%] | 89.5 | 91.54 | 91.93 | **92.32** |
| F1 score | 0.89 | 0.91 | 0.92 | **0.92** |
| $T_2$: Event position estimation | | | | |
| RMSE [m] | 3.31 | 3.43 | 2.85 | **2.73** |
| MAE [m] | 2.36 | 2.46 | 1.84 | **1.66** |
| $T_3$: Event characterization | | | | |
| ($RMSE_R$, $RMSE_L$) | (10.38, 1.48) | (6.33, 1.03) | (6.15, 1.1) | **(4.76, 0.8)** |
| ($SMAPE_R$, $SMAPE_L$) | (57.43, 60.94) | (54.59, 55.44) | (55.07, 54) | **(53.86, 54)** |
| $T_4$: Event identification | | | | |
| Accuracy [%] | 88 | 89 | 90 | **91** |
| AUC | 0.92 | 0.93 | 0.93 | **0.94** |

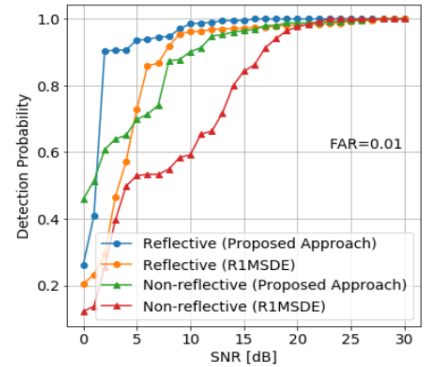

Fig. 5: Our model vs. conventional method comparison

The proposed framework is compared with a conventional rank-1 matched subspace detector (R1MSDE [4]) using an unseen test dataset including reflective, non-reflective and no event sequences. The results depicted in Fig. 5 show that our approach achieves a notable improvement over the conventional technique particularly for low SNR values.

### 4. Conclusion

We proposed a multi-task framework based on BiLSTM-CNN for fiber fault diagnosis by processing noisy OTDR data. The experimental results demonstrated that our approach significantly outperforms different ML methods as well as the conventional OTDR event analysis technique.

The work has been partially funded by the German Ministry of Education and Research in the project OptiCON (#16KIS0989K).